\documentstyle[epsf,graphicx]{mn}

\def\x2{$\chi^{2}$}

\def\ginga{{\it Ginga}}
\def\exosat{{\it EXOSAT }}

\newbox\grsign \setbox\grsign=\hbox{$>$} \newdimen\grdimen \grdimen=\ht\grsign
\newbox\simlessbox \newbox\simgreatbox \newbox\simpropbox
\setbox\simgreatbox=\hbox{\raise.5ex\hbox{$>$}\llap
     {\lower.5ex\hbox{$\sim$}}}\ht1=\grdimen\dp1=0pt
\setbox\simlessbox=\hbox{\raise.5ex\hbox{$<$}\llap
     {\lower.5ex\hbox{$\sim$}}}\ht2=\grdimen\dp2=0pt
\setbox\simpropbox=\hbox{\raise.5ex\hbox{$\propto$}\llap
     {\lower.5ex\hbox{$\sim$}}}\ht2=\grdimen\dp2=0pt

\def\asca{{\it ASCA~}}

\def\ginga{{\it Ginga~}}
\def\exosat{{\it EXOSAT~}}

\begin{document}

\title[{\it BeppoSAX} Observations of Seyfert-2 Galaxies]
{{\it BeppoSAX} Observations of the Seyfert-2 Galaxies NGC 7172
and ESO 103-G35}

\author[Akylas et al.]
{\Large A. Akylas$^{1,2}$,  I. Georgantopoulos$^{1}$, A. Comastri$^{3}$     \\
$^1$ Institute of Astronomy \& Astrophysics, National Observatory of Athens, I. Metaxa
$\&$ B. Pavlou, Penteli, 15236, Athens, Greece \\
$^2$ Physics Department University of Athens, Panepistimiopolis, 
 Zografos, 15783,
Athens, Greece \\  
$^3$ Osservatorio Astronomico di Bologna, Via Ranzani 1,
I-40127 Bologna, Italy
}

\maketitle

\begin{abstract}

We investigate the X-ray spectra of the type-2 Seyfert galaxies
 NGC 7172 and ESO 103-G35, 
using {\it BeppoSAX} observations, separated by approximately one year. 
We find that the X-ray spectra of both  NGC 7172 and ESO 103-G35 can be  
well fitted using a power-law model with an  Fe K$_\alpha$ emission line at 6.4 keV. 
We did not find any statistically significant
evidence for the existence of a reflection component in the X-ray spectra 
of these two galaxies.  
 The continuum flux has decreased by a 
factor of approximately two during this period, in  both objects.
 However, the spectral index as well as the 
 absorption column have remained constant. 
 We find weak  evidence for the decrease of the normalization 
of the Fe K$_\alpha$ emission line in a similar manner 
 to the continuum in NGC7172. We also report evidence for a broad  
Fe K$_\alpha$ confirming previous \asca observations.    
In contrast, in the case of ESO 103-G35 the line flux 
does not change while its width remains unresolved.

\end{abstract}


\newpage

\section{Introduction}

In recent years X-ray missions such as \exosat \ginga and  \asca  have 
 demonstrated that Seyfert-2 galaxies have a very complex 
 X-ray spectrum (see Mushotzky, Pounds \& Done 1993 for
 a recent review). 
 In particular their spectrum contains at least the following features:
a) a power-law continuum with slope $\Gamma \sim 1.9$
 absorbed by a very large column ($>10^{23} \rm cm^{-2}$)
 (eg Turner \& Pounds 1988); this obscuring screen is probably associated with 
 a molecular torus.   
 b) an FeK line  at 6.4 keV ie originating from neutral Fe 
 c) a spectral curvature (hump) above 10 keV; 
 this is generally believed to 
 originate from reflection of the power-law component
 on  cold material (eg Lightman \& White 1988, George \& Fabian 1991).

 In contrast to the recent progress in understanding 
 the spectral features of Seyfert-2 galaxies, our 
 knowledge of their X-ray variability properties remains scanty. 
 Usually, this has been based on 
 comparison of the X-ray flux between observations from 
 different missions spanning time intervals of several years.
 Recently, Turner et al. (1997) have detected 
 some evidence for  short term 
 variability in a handful of Seyfert-2 galaxies using \asca.
 The systematic monitoring of Seyfert-2 galaxies has become possible 
 with the {\it RXTE} mission. {\it RXTE} observations of a few objects 
 revealed strong variability in time-scales of hours as 
 well as evidence for spectral variability (Georgantopoulos et al. 1999, 
 Georgantopoulos \& Papadakis 2000).    
 Indeed, time variability studies can provide strong constraints 
 in understanding the geometry of the circumnuclear matter 
 in these objects. 
 For example, 
 the time lags between the power-law
 continuum and the normalization of the  Fe line could constrain  
 the origin of the line. In the case where the 
 Fe line originates in an accretion disk (as is the standard 
 scenario for Seyfert-1 galaxies), the Fe line should 
 closely track the variations of the intrinsic power-law.

In this paper we present an analysis of four observations of the  
Seyfert-2 galaxies NGC 7172 and ESO 103-G35 from the 
{\it BeppoSAX} archive. The two observations   for each galaxy 
 are separated by a year's interval. 
Our goal is to study any long-term variations in the spectral 
features between the two epochs. 
The broad energy band of $\it BeppoSAX$ as well as its good spectral resolution, 
 provide the opportunity to place stringent constraints 
 on the geometry of the nuclei of ESO103-G35 and NGC7172 and 
 any surrounding gaseous media.

\section{The sample}

\subsection{NGC 7172}
\begin{figure}
\rotatebox{0}{\includegraphics[height=9.0cm]{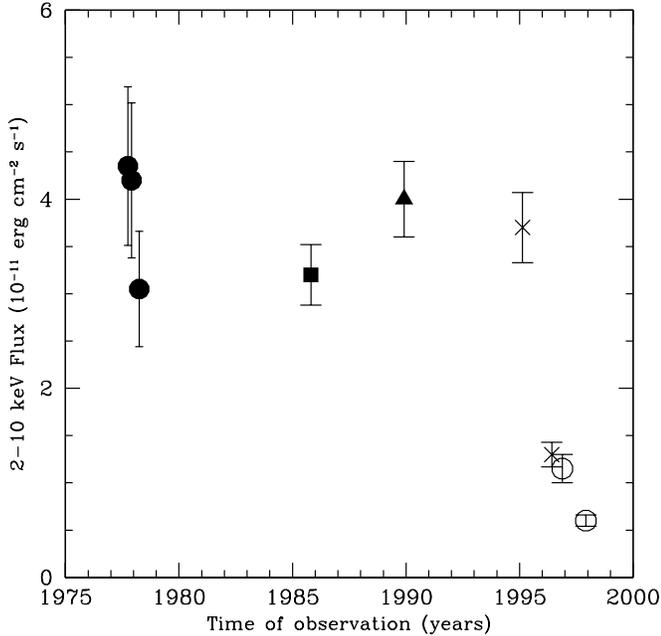}}   
\caption{Long term X-ray variability curve for the 2-10 keV absorbed continuum flux of 
NGC 7172 including $\it HEAO-1$ (filled circles), $\it EXOSAT$ (filled squares), 
$\it Ginga$ (filled triangles), $\it ASCA$ (crosses), and $\it BeppoSAX$ (open circles) 
observations.}  
\label{ngc7172_history}
\end{figure}

NGC 7172 is an obscured, almost edge on galaxy. It belongs to the compact group
HCG90 and  has a redshift of z=0.0086. It is classified as 
a type 2 Seyfert galaxy. Strong infrared emission from the nucleus
is observed, which is variable on time scales of approximately 3 months  
(Sharples et al. 1984). The first X-ray observation which was performed 
by $\it EXOSAT$ showed a $\Gamma$=1.84 power-law absorbed 
 by $N_H \sim 10^{23} \rm cm^{-2} $ 
 (Turner \& Pounds 1989). Later observations with $\it Ginga $ confirmed  the 
above results and also indicated that a reflection component may be important
 (Smith \& Done 1996). Furthermore the analysis of
 \asca observations (Turner et al. 1997) showed the presence of a 6.4 keV
emission line with an equivalent width (EW) of $68^{+36}_{-35}$ eV. In 
contrast with the previous results they found a rather harder photon index,   
$\Gamma\sim1.7$, while the amount of the absorption column remained constant.  
Guainazzi et al. (1998), using 2 $\it ASCA $  observations separated by one year,
found an even flatter photon index of $\Gamma \sim 1.5$ absorbed by the 
same  column density found in previous observations.
 They further detect short-term  variability (within a day) of about 30 percent. 
 Until 1995, before the {\it BeppoSAX} observations, the flux of NGC 7172 has remained  
almost constant between $\sim$ 3 and $\sim5 \times 10^{-11} \rm ~erg~cm^{-2}~s^{-1}$.
 Since May 1996 the source has reached a very low flux state of 
$\sim 1 \times 10^{-11} \rm ~erg~cm^{-2}~s^{-1}$. 
The flux variability  history of NGC7172 is summarized in Fig. \ref{ngc7172_history}.

\subsection{ESO 103-G35}
\begin{figure}
\rotatebox{0}{\includegraphics[height=9.0cm]{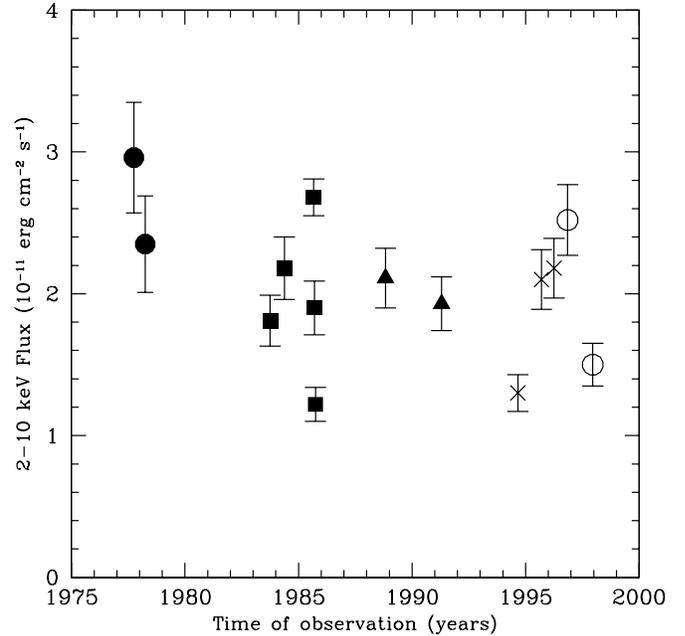}}   
\caption{Long term X-ray variability curve for the 2-10 keV absorbed continuum flux of 
ESO 103-G35 including $\it HEAO-1$ (filled circles), $\it EXOSAT$ (filled squares), 
$\it Ginga$ (filled triangles), $\it ASCA$ (crosses), and $\it BeppoSAX$ (open circles) 
observations.} 
\label{eso103_history} 
\end{figure}

The HEAO A2 hard X-ray source 1H 1832-653 was identified as the S0/Sa
type galaxy ESO 103-G35, with a  redshift of z=0.013, by Phillips et al. (1979).
They classified the AGN as a type 1.9 Seyfert galaxy because of the 
weakness of $H\beta $ emission line and the presence of a broad $H\alpha$ line.
$\it EXOSAT$ observations, showed a steep
photon index of $\Gamma=1.92^{+0.34}_{-0.34}$  absorbed by a high column 
density of  N$_H>10^{23}~ \rm cm^{-2}$, which was found to vary within factor of 
2 in a period of 90 days possibly due to motion of material in the Broad 
 Line Region  across 
the line of sight (Warwick, Pounds \& Turner 1988). 
Later observations obtained by $\it Ginga $ strongly suggested the presence of a 
narrow emission line at 6.4 keV and also of a  reflection component with a  photon 
index $\Gamma=2.19^{+0.08}_{-0.08}$ (Smith \& Done 1996).
$\it ASCA $ observations (Turner  et al. 1997) confirmed the previous 
$\it EXOSAT $ results and showed strong evidence for the presence 
of a triplet Fe line (6.4, 6.68, 6.98 keV). They also found a decrease by a factor of 
 two in the continuum flux as compared to $\it Ginga $. Finally, Forster,Leighly 
\& Kay (1999), analyzing three $\it ASCA $ observations spanning a period of $\sim$2 
years (1994-1996), found a decrease in the EW of the Fe line while the continuum 
flux increased by a factor of 2. An edge at 7.4 keV was also found in the 1996 
observation.  
The long term variability  history of ESO 103-G35 is summarized in Fig.  
\ref{eso103_history}.

\section{Observations}
The scientific instrumentation on board the Italian-Dutch X-ray  Satellite 
{\it BeppoSAX} includes a Medium Energy Concentrator Spectrometer, MECS, 
which consists of three units, (Boella et al 1997) 
a Low Energy Concentrator Spectrometer, LECS, (Parmar et al 1997) 
a High Pressure 
Gas Scintillation Proportional Counter, HPGSPC, (Manzo et al 1997)
and a Phoswich Detector System, PDS, (Frontera et al 1997), all of which 
 point in the same direction.
The MECS instrument consists of a mirror unit plus a gas scintillation 
proportional counter and has imaging capabilities. It covers the energy range 
between 1-10 keV with a spatial resolution of about 1.4 arcmin at 6 keV and a spectral 
resolution of 8 per cent at 6 keV. The LECS instrument is similar to the MECS
and operates in the 0.1-10 keV band with the same spectral and spatial 
resolution of the LECS. The PDS is a direct view detector with 
rocking collimators and extends the $\it BeppoSAX$ 
bandpass  to high energies (13-300 keV). Its  energy resolution is 15 per cent at 60 keV. 
 In our analysis we use MECS and PDS data only. 
  Our objects have not been detected by the HPGSPC detector. 
  We did not use the LECS data as there are nearby 
  contaminating soft X-ray  sources, most probably stars in case of ESO 103-G35, while 
 in case of NGC 7172 the contamination is due to the presence of the Hickson compact 
 group H90. 
  These sources are not present in the hard MECS images.  
  Another reason for excluding the LECS data was the poor photon statistics 
  (see tables 1 and 2).

 We analyze here two observations of NGC7172 and  ESO 103-G35
 both  obtained with $\it BeppoSAX$  within an interval of about one year.
 More specifically, NGC7172 was first observed in 1996 (from October 15th  to 16th)
and then in 1997 (from November 6th to 7th). ESO 103-G35 
was first observed in 1996 (from October 3rd to 4th)
and then in 1997 (from October 14th to 15th). 
Tables 1 and 2 list the exposures 
times and net count rates (background subtracted) for all four 
observations. 
The three different (or two after May 7 1997, when MECS1 failed) MECS units
are merged together in order to increase the signal to noise ratio. This is 
feasible because the three MECS units show very similar performance and  
the difference in the position of the optical axis in the three units is 
smaller than the scale on which the vignetting of the telescopes varies 
significantly ($>$5 arcmin).

\begin{table}
\begin{center}
\caption{{\it BeppoSAX} observations of NGC 7172}
\begin{tabular}{ccccccc}

Instrument  &  Exposure time (sec)  &  count rate  \\
\hline 
15/16-October-1996 \\
\hline
LECS &  15490      & $0.038\pm 0.002$      \\
MECS(1,2,3) &  39172 & $ 0.148 \pm 0.002 $  &         \\
PDS  &  17277 & $ 0.431 \pm 0.051 $ &        \\
\hline      
6/7-November-1997  \\
\hline
LECS &  22896 &  $0.022\pm 0.001$ \\
MECS(1,2) &  49437 & $ 0.054 \pm 0.001 $ &         \\
PDS  &  21146 & $ 0.292 \pm 0.047 $ &         \\
\hline

\end{tabular}
\end{center}
\end{table}     

\begin{table}
\begin{center}
\caption{{\it BeppoSAX} observations of ESO 103-G35}
\begin{tabular}{ccccccc}

Instrument & Exposure time (sec) & count rate \\
\hline 
3/4-October-1996        \\
\hline
LECS &  10238 &  $0.076\pm 0.003$             \\     
MECS(1,2,3) &  50619 &   $0.308 \pm 0.003$    &           \\
PDS  &  21029 &   $0.746 \pm 0.047 $  &          \\
\hline
14/15-October-1997         \\
\hline
LECS &   3699  &    $0.04\pm 0.004$             \\
MECS(1,2) &  14347 &  $ 0.123 \pm 0.003$         &       \\
PDS  &   5915 &   $0.579 \pm 0.087$         &      \\
\hline
\end{tabular}
\end{center}
\end{table}

\section{Timing Analysis}

\begin{figure}
\rotatebox{0}{\includegraphics[height=9.0cm]{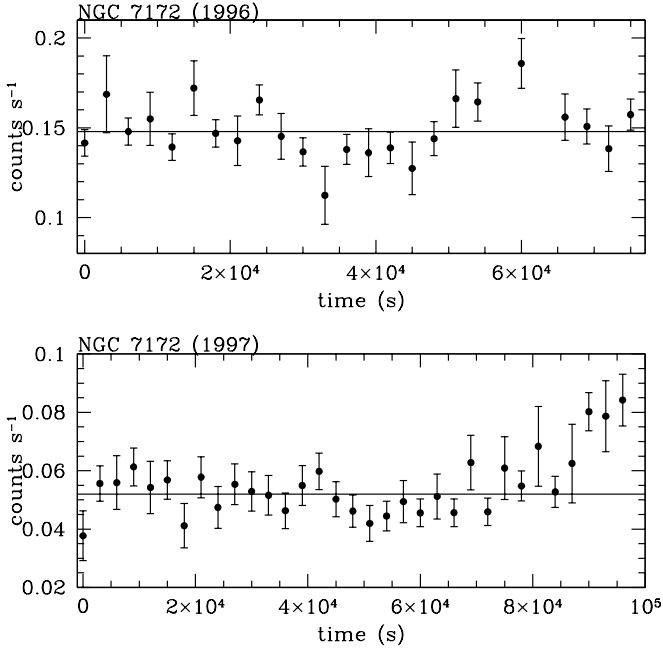}}   
\caption{ Background subtracted Light curves of NGC 7172 observations (1996 upper panel and 1997 bottom panel)
for the MECS instrument (1.65-10 keV). 
The binning time is 3 ks.}
\label{7172curve} 
\end{figure}

\begin{figure}
\rotatebox{0}{\includegraphics[height=9.0cm]{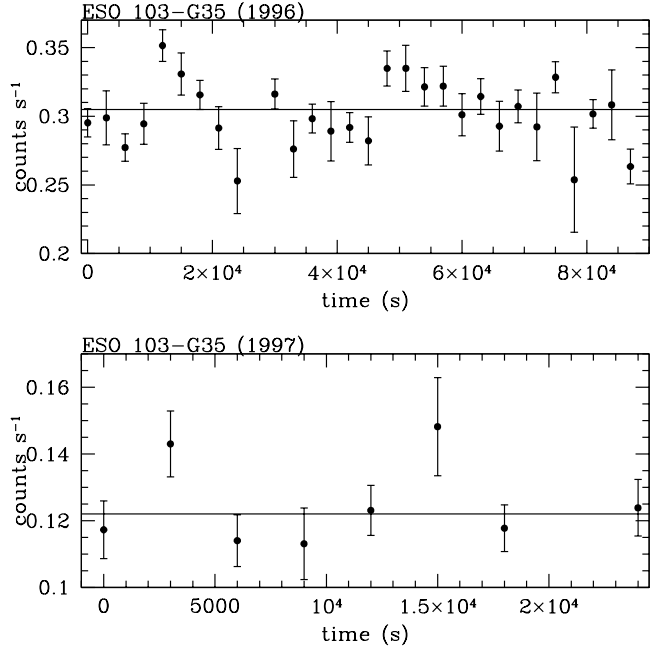}}   
\caption{ Light curves (background subtracted) of ESO 103-G35 observations (1996 upper panel and 1997 bottom 
panel) for the MECS instrument (1.65-10 keV). 
The binning time is 3 ks}
\label{esocurve} 
\end{figure}

In Fig. \ref{7172curve} and \ref{esocurve}
  we present the light curves for 
the MECS instrument for each observation separately,
using a binning time of 3 ks 
 (the errors correspond to the 68 per cent confidence level).
 When we perform  a constant fit, the resulting
 \x2 implies short term variability in both objects. 
In particular, for the first observation of NGC 7172 we find 
that the best fit average count rate ($0.147 \pm 0.007$) gives a \x2=36 
(for 24 degrees of freedom, dof),
while for the second observation we find  
$0.052 \pm 0.005$ with \x2/dof=64/33.  
In the case of ESO 103-G35 the best fit constant is
 $0.305 \pm 0.004$ corresponding to  \x2/dof=72/29 for the 
first observation while for the second one we find 
$0.122 \pm 0.004$ with \x2/dof=10/8. 
The above \x2 suggest the presence of statistically significant ($>3\sigma$)
 short-term variability  in the 
  1997 and 1996 observations of NGC7172 and ESO103-G35 respectively. 
  Our results are compatible with  previous \asca findings. 
More specifically, Guainazzi et al (1998) detected 
a flux variability of about 30 per cent, on time scales of hours
 in a 1996 $\it ASCA$  observation of NGC 7172.   
Furthermore, Forster et al (1999) using $\it ASCA$ data  found
statistically significant short term variability  in a 1996 observation of 
ESO 103-G35. 

The variability amplitude was estimated in all four observations by means of 
the excess variance $\sigma^{2}_{rms}$ (for a definition of this quantity see 
Nandra et al 1997). In table 3 we present the values of $\sigma^{2}_{rms}$
and the unobscured luminosity for each observation.    
Unfortunately a comparison with the results  found by Nandra et
 al (1997) for a sample of Seyfert 1 galaxies is not straightforward. This is  
due to both the  different exposure times and binning time we have used. 
 For example our large binning time has
the effect to suppress the $\sigma^{2}_{rms}$ value. Indeed our values are 
slightly lower than the values quoted in Nandra et al (see their Fig 4).

\begin{table}
\begin{center}
\caption{The values of the variability amplitude, $\sigma^{2}_{rms}$,
 and the unobscured luminosity for each observation}
\begin{tabular}{ccccccc}

object & date & $\sigma^{2}_{rms}$ & luminosity (ergs s$^{-1}$) \\
       &      &               & 2-10 keV \\

\hline 
NGC 7172 & 1996 & 0.0045  & $3.5\times 10^{42}$          \\     
NGC 7172 & 1997 & 0.0018    & $1.7\times 10^{42}$          \\     
ESO 103-G35 & 1996 & 0.0029  &  $2.7\times 10^{43}$          \\     
ESO 103-G35  & 1997 & 0.004  &   $1.5\times10^{43}$          \\     
\hline
\end{tabular}
\end{center}
\end{table}

\section{Spectral Analysis} 
We used an extraction 
 radius of four arcmin. The above area encircles more than  
85 per cent of the photons in the 2-10 keV energy band. 
The spectrum of the  background was estimated from  source free regions of the
image. 
In particular, we consider an annulus with inner radius 6 arcmin and outer radius 12, 
for the MECS instrument. 
 We used the Rise Time threshold background subtracted PDS 
spectra released from  the {\it BeppoSAX} archive. 
We use data between the energy ranges 1.65-10 keV  and 15-50 keV
for the MECS and PDS detectors respectively 
 where  the response matrices are well calibrated.
We do not consider the PDS data above 50 keV as in this energy range the signal to noise
ratio is low.
The MECS spectral files were rebinned  linearly to give a minimum of 20 counts 
per channel while the PDS spectral files were binned logarithmically in 18 channels.
The spectral fitting was carried out using the {\sl XSPEC} v10 software package.
 All errors quoted in the best-fit spectral parameters correspond to the 
 90 per cent confidence level.  
The MECS and PDS data were fitted simultaneously. A relative normalization factor 
was introduced between the PDS and MECS data spectrum since the {\it BeppoSAX} instruments 
show some mismatches in the absolute flux calibration.
 We assumed a PDS to MECS normalization factor between 0.77 to 0.95 
 (Fiore, Guainazzi \& Grandi 1999).

\subsection{NGC 7172 (1996)}

We first fit a simple model consisting of a  power-law absorbed 
 by a neutral column $N_H$ ( Fig. \ref{ngc7172}).    
This provides a good fit to the data ($\Delta \chi^2$=145 for 146 dof).  
The addition of a broad Gaussian line improves significantly the fit 
($\Delta \chi^2$=10). 
The best-fit line energy was $6.0^{+0.6}_{-0.6}$ keV. 
Since this value is consistent with 6.4 keV, which indicates a neutral 
or weakly ionized Fe, (e.g. Nandra et al 1997) 
we fix it at 6.4 keV in the rest of the analysis.  
 The line width is $\sigma_{K_\alpha}=0.8^{+0.3}_{-0.4} $ keV, 
so it is  marginally resolved given the energy resolution of the MECS.
 When we fix the line width at 0.1 keV we obtain an increase of 
 $\Delta \chi^2 $=2.8. 
The best-fit parameters  are presented in table 4. 

In fig 5 residuals can be seen above 7 keV.  
 We therefore attempted to  
include an absorption edge at $\sim$7.2 keV.   
We found that the addition of the edge significantly improves the fit 
($\Delta \chi^2$=3.4 for the addition of one parameter). 
The optical depth of the edge is $0.18^{+0.11}_{-0.12}$.
The rather flat spectral index $\Gamma=1.64^{+0.12}_{-0.09}$ could suggest 
that a reflection continuum is present.
 Therefore, we further tried to add a reflection component to our model.
 We used the {\sl PEXRAV}  model 
in {\sl XSPEC} (Magdziarz \& Zdziarski 1995). 
 We assumed that both the reflection component and the power-law 
 are absorbed by the same column density. 
We fix the {\sl PEXRAV} normalization to that of the power-law.  
 We also arbitrarily fix the inclination angle 
(of the plane of the reflecting material) 
 to $i=60^\circ$, since the shape of the reflection spectrum 
 in our energy band is relatively independent of the inclination angle. 
 The cut-off energy of the incident power-law was set at 300 keV. Finally, 
 the Fe and light 
 element abundances were kept fixed at the solar abundance values. 
We find that a reflection component is not statistically significant 
($\Delta \chi^2 \sim 1.5$ for one additional 
parameter). However the rather uncertain  value of the reflection parameter 
(R$=1.2^{+0.7}_{-0.9}$) and the steeper spectral photon index of
$\Gamma=1.78^{+0.26}_{-0.19}$ ,which is closer to the canonical AGN spectral 
index (Nandra \& Pounds 1994) suggest that a 
reflection component cannot be ruled out. The fraction of the intrinsic
reflected flux relative to the total in the 2-50 keV band is 0.25.

\subsection{NGC 7172 (1997)}

\begin{figure*}
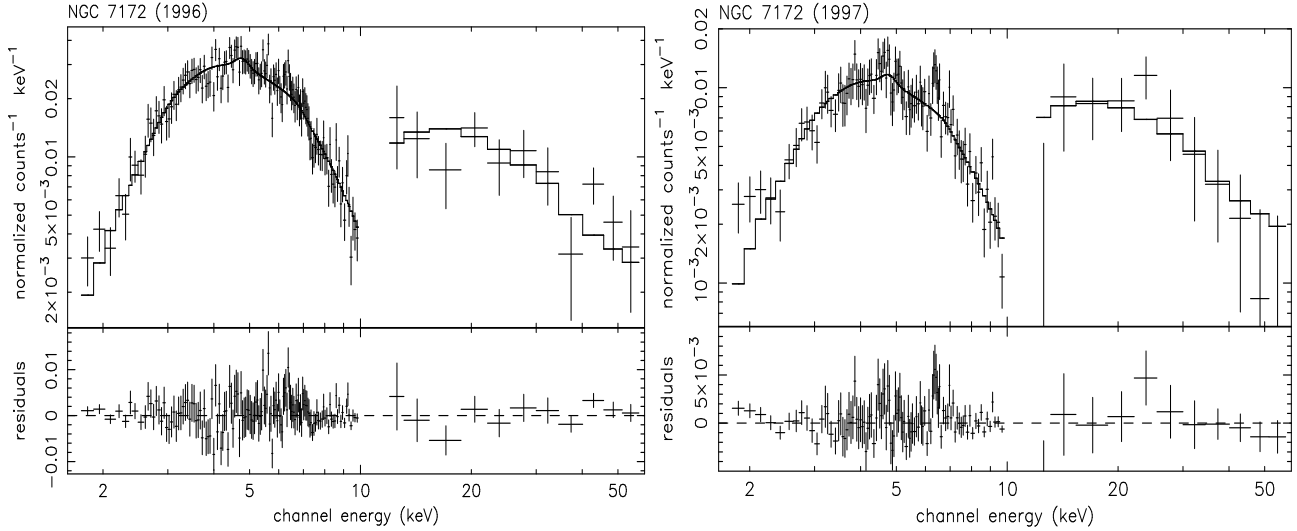

\hfill
\rotatebox{270}{\includegraphics[height=8.5cm,width=7.0cm]{res7172.ps}}   
\rotatebox{270}{\includegraphics[height=8.5cm,width=7.0cm]{res7172_r.ps}}   
\caption{Data, folded model (simple absorbed power-law model) and 
residuals  for both 1996 (left panel) and 1997 (right panel) observations of NGC 7172}
\label{ngc7172}
\end{figure*}

Similar to the 1996 observation of NGC 7172, an absorbed power-law model 
($\Gamma\sim 1.5$) provides  a good fit
($\Delta \chi^2$=112 for 109 dof, see  Fig. \ref{ngc7172}).  
The addition of a Gaussian line improves the fit significantly
($\Delta \chi^2=6)$.
The line energy  is $ 6.74^{+0.26}_{-0.42}$ keV
 consistent with neutral Fe and therefore  hereafter we fix the line energy 
at 6.4 keV. 
 The range of the line width is 0.08-1 keV.
The  best fit parameters  are shown in table 4. 
 Inspection of the residuals shows that there is no evidence for the presence 
of an absorption edge.
 However, we attempt to include an edge at 7.25 keV based on the previous 
1996 results. We found that there is no improvement in the fit 
($\Delta \chi^2<1$). The optical depth of the edge is $0.08^{+0.15}_{-0.08}$.  
The rather flat spectral index $\Gamma=1.53^{+0.08}_{-0.16}$ 
could again suggest the presence of a reflection component.
However, when we try to include such a component, we find that, although it does not 
improve the fit ($\Delta \chi^2=1.5$ for one additional parameter),  
the photon index becomes steeper $\Gamma=1.81^{+0.60}_{-0.40}$ and the 
reflection parameter is R=$2.5^{+1.3}_{-1.8}$. The unabsorbed 
reflected flux is 40 percent of the total flux in the 2-50 keV energy 
band.  Comparison between the 
 spectral fits in the 1996 and 1997 observations 
 reveals that there is no  change 
 in the photon index and the column density within the errors.  

Finally, we perform joint fits in order to investigate any possible 
variability in the normalization of the different spectral components. 
 We fit both NGC 7172 observations with an absorbed power law 
model ($\Gamma=1.62^{+0.07}_{-0.09}$ and N$_H=9.1^{+0.3}_{-0.6}\times10^{22}~
\rm cm^{-2}$)  
plus a Gaussian component at 6.4 keV to account for the Fe line. The width of 
the line is $0.6^{+0.4}_{-0.5}$ keV and therefore remains unconstrained.  
Comparison of the two observations reveals  weak evidence for  
 spectral variability.  
The power-law normalization has decreased by about a factor of two 
between the 1996 and 1997 observations. 
 This may be followed by a similar change in the Fe line normalization. 
 This is evident in Fig. \ref{ngc7172_contours} where we present the 68, 90 
and 99 per cent confidence contours for the Fe line 
 against the power-law normalization. 
 The contours were extracted using the above best fit model. The EW 
  are $280^{+140}_{-130}$ and $233^{+190}_{-160}$ eV for the 1996 and 1997 
observations respectively. However we note that when we fix the width of 
the line at 0.1 keV (this increases \x2 by 2.5)  
we do not detect any significant variability in the normalization of the 
line.

\subsection{ESO 103-G35 (1996)}

A poor fit  (\x2=240 for 173 dof) is obtained with a single power-law model,
   emphasizing the importance of an Fe
emission line. This is again evident in  Fig. \ref{eso103} where we plot the 
absorbed power-law model and  residuals.   
Indeed when we include a  Gaussian line  the \x2 is reduced by 66. 
The Fe line has an energy of $6.37^{+0.09}_{-0.10}$ keV 
 consistent with neutral Fe  and therefore hereafter we fix it at 6.4 keV.
 Its width is $\sigma=0.3^{+0.1}_{-0.1}$ keV and therefore the line remains  
essentially unconstrained.  
The photon index, $\Gamma=1.87^{+0.06}_{-0.09}$, is consistent with the 
typical value of 1.9 for Seyfert galaxies (Nandra \& Pounds 1994).
The best-fit spectral parameters are presented in table 5.
We further investigate the  existence of excess Fe absorption near 7.5 keV
 following the claims of Warwick et al. (1993). 
 Although there is some hint for an Fe edge in Fig. \ref{eso103}, 
 we find no significant improvement when we include an 
edge component ($\Delta \chi^2=1.5$ for two additional 
parameters). The edge is found at $7.35^{+0.65}_{-0.35}$ keV with  optical 
depth $0.1^{+0.08}_{-0.10}$.
 Finally, based on the $\it ASCA $ results (Turner et al 1997) we tried to 
include two 
emission lines at 6.68 and 6.96 keV (ionized Fe) but the reduction in 
\x2 was less than 1. A reflection component is not statistically significant as 
the reduction of \x2 is very low ($\Delta \chi^2$= 1.1 for  one additional parameter). 
Note that the power law spectral index in the 
case of a reflection component is  $\Gamma=2.0^{+0.2}_{-0.2}$ while 
R=$1.2^{+0.4}_{-0.6}$. The intrinsic flux of the reflection component corresponds to 
the 20 per cent of the total flux in the 2-50 keV energy band.

\subsection{ESO 103-G35 (1997)}
 
An acceptable fit is obtained 
when we use the  absorbed power-law model  with \x2=70 for 71 dof,
 (see  Fig. \ref{eso103}). 
 An FeK emission line is highly significant yielding  
 $\Delta \chi^2$=14. 
The Fe line energy is  $6.5^{+0.10}_{-0.15}$ keV and therefore 
 again we fix it at 6.4 keV. 
 The line is unresolved: its width range is 0-0.4 keV.
 The photon index, $\Gamma=1.81^{+0.09}_{-0.30}$, is consistent with the 
 previous 1996 observation. The results are presented in table 5.  
  A reflection component cannot be ruled out as the reflection parameter
is R=$1.6^{+2.0}_{-1.2}$. However,the reduction in \x2 is very low 
($\Delta \chi^2$= 1.7 for one additional parameter). When the reflection component is 
 included the power law spectral index steepens to $\Gamma=1.97^{+0.30}_{-0.30}$. 
The ratio between the unabsorded reflected flux and the total flux in the 
2-50 keV energy band is 0.3.  
Next, we investigate the  existence of an Fe edge at 7.35 keV.
However, the reduction in \x2 was less than 1 for one  
additional parameter. 
The optical depth of the edge is $0.11^{+0.10}_{-0.11}$.
In general, we detect no spectral variability between the
 1996 and 1997 observations in the sense that both the 
absorption column and the photon index remain constant.

We perform a joint fit in the 1996 and 1997 observations in order to detect any 
possible line flux variations.
We used a model consisting of an absorbed power-law plus a Gaussian line at 
6.4 keV. The best-fit photon index is $1.83^{+0.09}_{-0.09}$ and the column density
is $18.5^{+0.2}_{-0.8}\times10^{22}~\rm cm^{-2}$. The line is unconstrained, ie the range of the line width 
 is 0.15-0.36 keV. The EW are $220^{+120}_{-50}$ 
and $330^{+170}_{-140}$ eV for the 1996 and 1997  observations respectively.
 The continuum flux is reduced approximately by a factor of two.
 However,  there is no evidence  that this  variation is followed by the line flux, in 
contrast to the case of NGC7172
(see Fig. \ref{eso103_contours}  where we plot the 68, 90 and 99 per cent 
joint contours for the power-law and Fe line normalization).

 \begin{table*} 
\tabcolsep 3pt
\caption{Best fit parameters for the two observations of NGC 7172}
\begin{tabular}{ccccccccc}

Date & N$_{H}$ & $\Gamma$ & $A_{pl}$ & $\rm Energy$ & $\sigma$ & $A\rm _{ga}$ & 
$\rm EW$ & \x2/dof  \\
   & $(10^{22}~\rm cm^{-2})$ &   &$(10^{-3}~\rm cts~s^{-1}~keV^{-1})$   
 & $\rm (keV)$ & $\rm (keV)$ &$(10^{-5}~\rm cts~s^{-1})$  & $\rm (eV)$ &    \\
\hline 
{15-10-96} & $9.0^{+0.9}_{-0.7}$ & $1.64^{+0.12}_{-0.09}$ &  
$4.1^{+1.2}_{-1.1}$ &  $6.4$ & 
$0.8^{+0.3}_{-0.4}$ & $5.6^{+2.5}_{-2.4}$ &  $280^{+210}_{-160}$ & 135/144  \\
{6-11-97} & $8.3^{+0.70}_{-0.65}$ & $1.53^{+0.08}_{-0.16}$ & 
$1.7^{+0.5}_{-0.5}$ & $6.4$ & $0.32^{+0.68}_{-0.24}$ & 
$1.5^{+1.1}_{-1.0}$ & $175^{+195}_{-140}$& 106/107   \\
\hline

\end{tabular}
\end{table*}

\begin{figure}
\rotatebox{270}{\includegraphics[height=6.0cm]{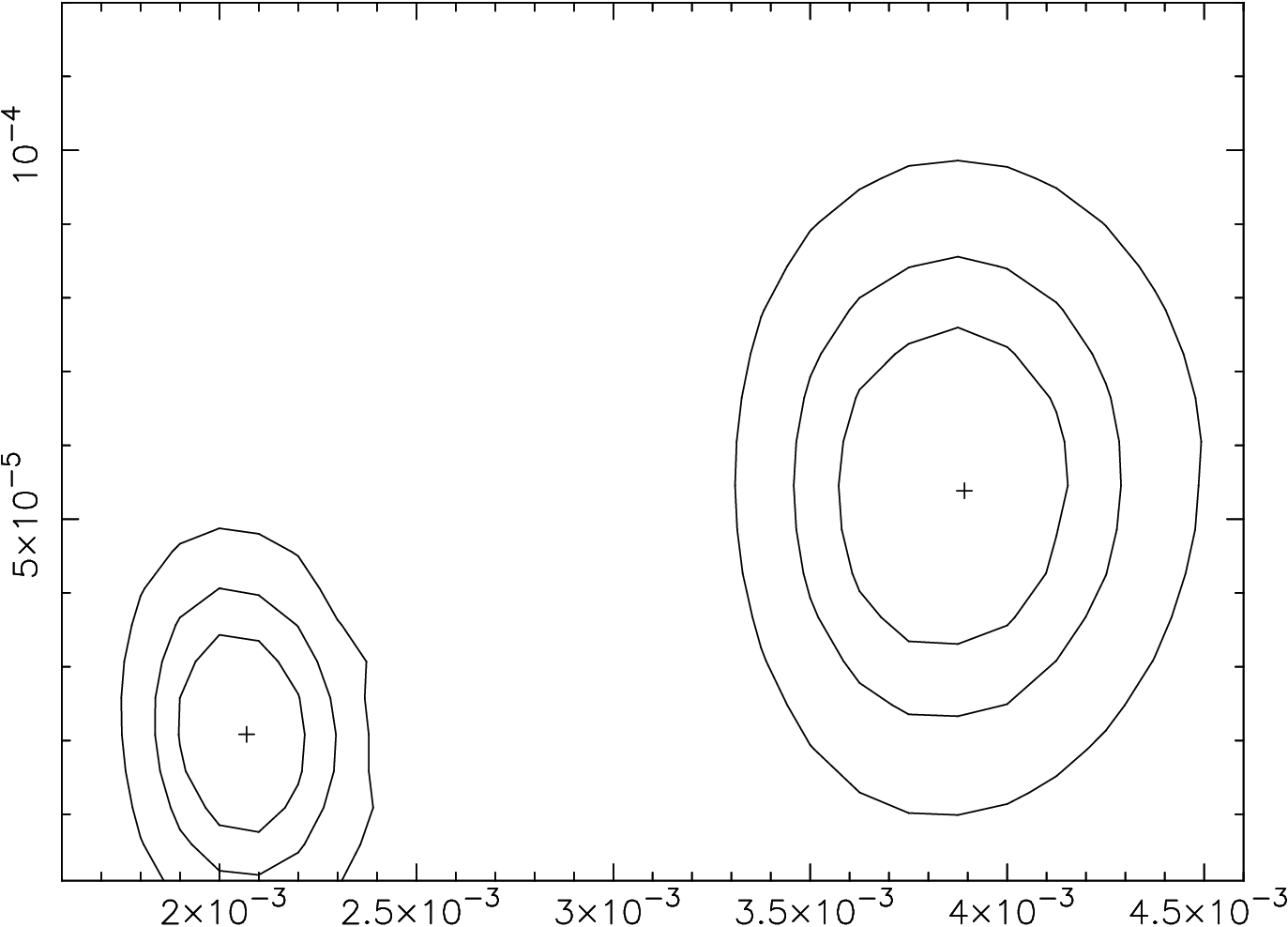}}   
\caption{Fe line flux vs power-law normalization contours for the 1996 and 1997 
observations of NGC 7172 in the joint fits. Confidence levels are 68, 90 and 99 per cent for two 
interesting parameters}
\label{ngc7172_contours}
\end{figure}

\begin{table*}
\tabcolsep 3pt 
\caption{Best fit parameters for the two observations of ESO 103-G35}
\begin{tabular}{ccccccccc}

Date & N$_{H}$ & $\Gamma$ & $\rm A_{pl}$ & $\rm Energy$ & $\sigma$ & $\rm A_{ga}$ &  
$EW $ & \x2/dof \\
   & $(10^{22}~\rm cm^{-2})$ & & $(10^{-3}\rm ~cts~s^{-1}~keV^{-1})$ & $\rm (keV)$ & 
$\rm keV$ & 
$(10^{-5}~\rm cts~s^{-1})$ &  $\rm eV $ &  \\
\hline 
{3-10-96} & $18.6^{+0.9}_{-1.0}$ & $1.87^{+0.06}_{-0.09}$ &  
$18.5^{+2.7}_{-3.3}$ &  $6.4$  &
$0.3^{+0.12}_{-0.12}$ & $15^{+5.0}_{-4.0}$ & $265^{+150}_{-100} $ & 174/171  \\
{14-10-97} & $19.1^{+1.4}_{-3.4}$ & $1.81^{+0.09}_{-0.30}$ & 
$9.6^{+2.6}_{-4.5}$ & $6.4$ & $0.16^{+0.24}_{-0.16}$ & 
$10.5^{+7.0}_{-5.4}$ & $290^{+500}_{-90}$ & 56/69 \\
\hline
\end{tabular}

\end{table*}

\begin{figure*}
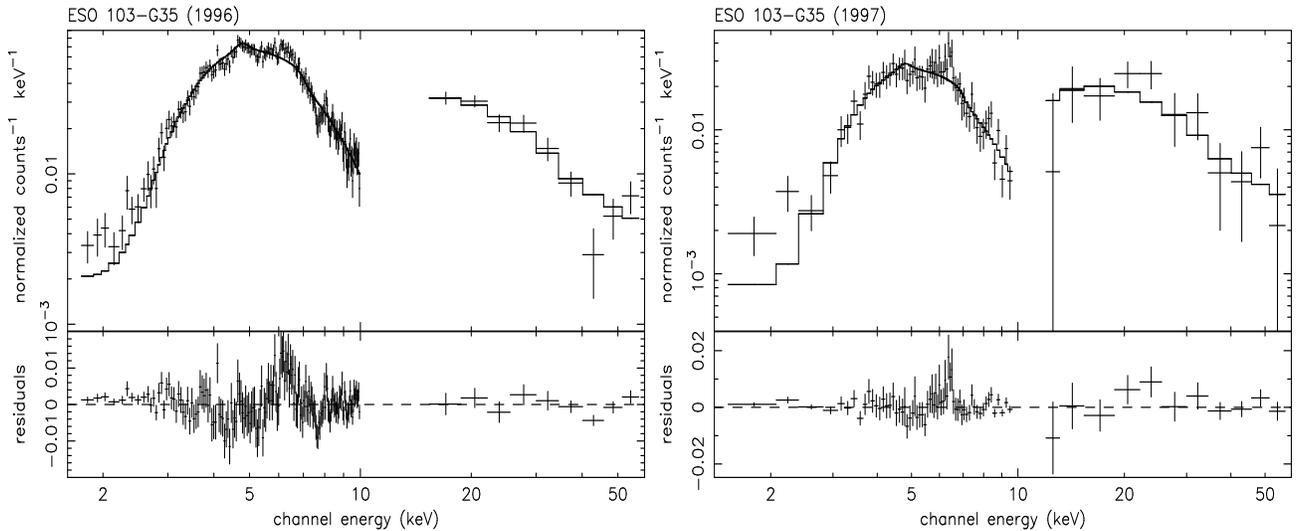

\hfill
\rotatebox{270}{\includegraphics[height=8.5cm,width=7.0cm]{esores.ps}}
\rotatebox{270}{\includegraphics[height=8.5cm,width=7.0cm]{eso_rres.ps}}
\caption{Data, folded model (simple absorbed power-law model) and 
residuals for both 
1996 (left panel) and 1997 (right panel) observations of ESO 103-G35} 
\label{eso103}
\end{figure*}

\begin{figure}
\rotatebox{270}{\includegraphics[height=6.0cm]{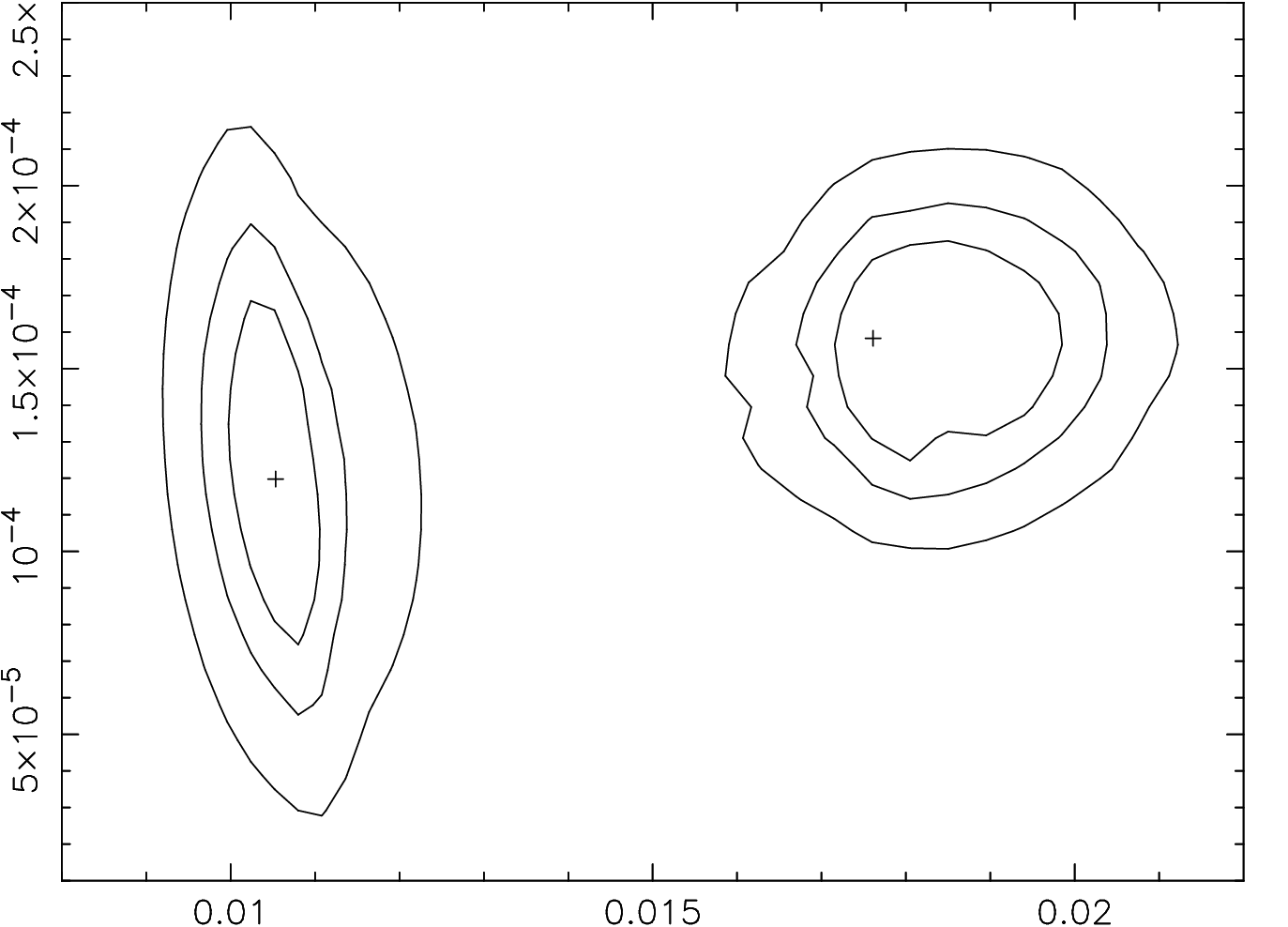}}   
\caption{Fe line flux vs power-law normalization contours for the 1996 and 1997 
observations of ESO 103-G35 in the joint fits. Confidence levels are 68, 90 and 99 per cent for two interesting parameters}
\label{eso103_contours}
\end{figure}

\section{discussion}

\subsection{NGC 7172}

The spectral analysis of both observations implies that NGC 7172 is 
absorbed by a column density of N$_H=9.1^{+0.3}_{-0.6}\times 10^{22}~\rm cm^{-2}$, 
 in good agreement with previous 
$\it EXOSAT$ (Turner \& Pounds 1989), $\it Ginga$ (Smith \& Done 1996)
and $\it ASCA $ (Turner et al 1997) observations.  
The data require the presence of  an  Fe line at 6.4 keV. 
 In the 1996 {\it BeppoSAX} observation the line is marginally resolved, 
at less than the 2$\sigma$ confidence level,  
 while in the 1997 observation (where the photon statistics are poorer) 
 the line width is unconstrained. 
 We note that the line has been resolved in previous \asca observations 
 (Guainazzi et al. 1998). 
There is some  evidence (Fig. \ref{ngc7172_contours}) for a variation in the 
normalization of the line.
 This is in agreement with Guainnazi at al (1998) 
 who presented similar evidence for a variation of the line based on two
$\it ASCA$ observations (1995 and 1996).
The line variation in conjunction with the line width has 
 important implications for the geometry of the 
 circumnuclear matter in NGC7172. 
 Indeed, the Fe line must originate  close to the nucleus 
 possibly in the outer regions of an accretion disk. 
 This would simultaneously explain 
  the fact that the continuum flux variations are 
 tracked by changes in the line flux, and  the large width of the line. Indeed,
the line broadening may be due  to the large velocities in the disk
 similar to that witnessed in MCG-6-30-15 by Tanaka et al. (1995). The
 observed broadening in the case of NGC 7172 implies a  velocity of 
$\sim$70000 $\rm km~s^{-1}$.
 We tried to investigate further the above hypothesis including 
 a line from  a relativistic accretion disk (for an accretion disk viewed at an 
 intermediate inclination angle with solar abundances).
The fit was only marginally improved ($\Delta\chi^2 \sim 1$).
We note that Georgantopoulos \& Papadakis (2000) find no significant variation of the 
normalization of the line in NGC 7172 using several {\it RXTE} observations spanning 
a period of around a week. This would set a lower limit of about a light week to the 
size of the line emitting region.     
 
 The observed EW cannot be interpreted in terms of  
 transmission processes in the torus alone. Indeed Leahy \& Creighton (1993)
 using Monte Carlo simulations found  that a column density of 
 N$_H \sim 9 \times 10^{22}\rm~cm^{-2}$ could produce an Fe line 
 EW of only $\sim60$ eV. This independently  suggests that a reflection 
component may be present.
Indeed, in this case the EW of the line should be increased 
by $\sim$150 eV 
due to reflection on the accretion disk (Reynolds, Fabian \& Inoue 1995, George \& 
Fabian 1991).
Alternatively the observed EW may be explained on the basis of a high Fe abundance. 
Indeed, excess absorption from  Fe is present in the first observation of NGC 7172 
at $\sim 7.2$ keV, confirming earlier reports by Warwick et al. (1993); 
however, note that superior spectral resolution observations with the 
$\it ASCA $ SIS (Turner et al 1997) do not corroborate this result.
The edge energy is consistent with a neutral or low 
ionization material surrounding the source (Kallman \& McCray 1982). The optical depth 
of the edge corresponds to an absorption column of $22.5^{+15.3}_{-14.2}\times10^{22}~
\rm cm^{-2}$. 
This large column density may suggest an Fe abundance well above the solar value. 
If this is the case, the observed EW could
be explained by transmission processes alone.
Finally note that cases where a strong Fe line is detected although there 
is no strong evidence for a reflection component are not uncommon 
(eg Gilli et al 2000, Reynolds et al 1995).

 The power-law photon index has a rather flat value of $\sim 1.6$
 while it remains constant in the two {\it BeppoSAX} observations. 
 The value of the photon index  is 
 in agreement  with previous \asca observations (Guainazzi et al 1998, Ryde 
et al 1997) 
who found a photon index of $1.47\pm 0.15 $. However, it is clearly flatter 
than 
 that found in  $\it EXOSAT$ (Turner et al 1989) and $\it Ginga$ 
(Smith \& Done 1996) 
 ($\Gamma \sim 1.8$). Therefore the above may be suggesting long term 
 spectral variability (Ryde et al. 1997). 
  For example, assuming that the  X-ray emission comes from Comptonization
of UV photons from an accretion disk on a hot electron corona 
(Rybicki \& Lightman 1979), 
 a steepening of the spectral index may be attributed to a change 
 of the temperature or the optical depth of the corona.
 Of course it is difficult to make a detailed comparison 
 due to differences in the energy bandpass of the various instruments 
 as well as in the spectral analysis: for example the addition of a 
reflection  component would result in the steepening of the spectral index. 
  However, we note that there is no correlation between the flux level and 
  the photon index observed in the case of the NGC7172 observations in the 
 past  20 years. Finally we note that  
 despite the flat photon index found above, our data do not significantly 
 support the 
 existence of any reflection continuum. A recent analysis of 
 $\it RXTE$ 
 monitoring observations of NGC 7172 (Georgantopoulos \& Papadakis 2000)
 again shows no evidence for reflection component. 
 
\subsection{ESO 103-G35}

The spectral analysis of the two observations of ESO103-G35 implies that a 
simple power-law model absorbed by a column density of 
 N$_H \sim 2 \times 10^{23}~\rm  cm^{-2}$ plus a Gaussian line  at 6.4 keV
provides a good fit to our data.
{\it EXOSAT} observations of ESO103-G35 (Warwick et al. 1988)
 reveal a 50 per cent decrease in the column density over a 90 
 day period. This variation was interpreted in 
 terms of an X-ray absorbing screen composed of broad-line clouds 
 moving around the X-ray source. 
 Alternatively, the N$_H$ variation could be attributed to photoionization 
 of the neutral column as the flux increases; 
 in this case an ionized Fe edge should be detected. 
However, in our analysis we detect 
 no variation of the column within the errors.  
 An Fe line at 6.4 keV is strongly needed to fit our data. 
Unfortunately, the width of the line remains unconstrained. 
However, {\it ASCA } SIS observations with  better spectral resolution 
have probably resolved the line into three components  at 6.4, 6.68 and 6.96 
keV (Turner et al 1997). 
Monte Carlo simulations (Leahy \& Creighton 1993) show that a column density of 
N$_H \sim 2 \times 10^{23}~\rm  cm^{-2}$ could produce an Fe line EW
  of $\sim$170 eV. However, in the case of ESO 103-G35 
 where the line originates possibly far away 
 from the central region, the EW varies accordingly to the variations of the 
 continuum. Although  a comparison is not straightforward, 
it is possible that a large fraction of the line emission
  could originate from transmission processes.  
    
Some residuals around 7.5 keV (see Fig. \ref{eso103}) may 
indicate the presence of an  absorption edge due to ionized Fe. 
This was first suggested by $\it Ginga$ results (Warwick et al. 1993) and was
 confirmed later with $\it ASCA $ observations (Forster et al 1999).    
 The power-law photon index has a steep value of $\sim 1.85^{+0.05}_{-0.05}$ 
which is consistent with the typical value 1.9 for Seyfert galaxies 
(Nandra \& Pounds 1994).
 Despite our large energy bandpass we do not  
find strong evidence for a reflection continuum
in contrast to previous 
$\it Ginga$ (Smith \& Done 1996) and $\it RXTE$ (Georgantopoulos \& Papadakis 2000)
 observations who found  evidence 
 for such a component but in a smaller band (3-20 keV). Higher signal to noise 
observations are necessary in order to test the presence of such a component.  

\subsection{SPECTRAL VARIABILITY IN SEYFERT 2 GALAXIES}
 Previous studies of spectral variability in Seyfert-2 galaxies 
 have shown some interesting results. In particular, 
 Iwasawa et al (1994) and Griffiths et al. (1998) 
 analyzed non-simultaneous {\it Ginga},
{\it ASCA}, {\it ROSAT} and {\it BBXRT} observations to investigate the spectrum of the 
Seyfert-2 galaxy  Mrk 3 . Their analysis demonstrated that the Fe line does show some 
variability on time-scales of a few years. 
Systematic monitoring observation became feasible with the {\it RXTE } mission.  
In particular, 
Georgantopoulos et al. (1999), Georgantopoulos \& Papadakis (2000),
 Smith, Georgantopoulos \& Warwick (2000) present monitoring observations 
 of several Seyfert-2 galaxies (Mrk3, ESO 103-G35,IC 5063, NGC 4507, NGC 7172 \& Mrk 348)
 spanning time periods from about seven days 
 to seven months. In most cases, they detect 
 spectral variability in the sense 
 that the column density decreases with increasing flux. 
 However, they find no evidence for Fe line variability.  
Here instead, we find tentative evidence for line 
variability in NGC 7172 in a period of one year 
confirming previous {\it ASCA} results (Guainazzi et al 1998). 
To our knowledge this is one of the very few examples  of Fe line variability 
in Seyfert-2 galaxies. Our finding  requires at least one of the line emitting regions
has dimension not much more than about a light year. It would seem more likely 
that this scale size corresponds to an accretion disk or inner  cloud
structure rather than the inner extent of the putative molecular torus. Indeed, the 
molecular torus is probably a much greater structure. At least in the case of 
NGC 1068 observations of molecular hydrogen (Tacconi et al 1994) showed that the 
obscuring torus is located between 30 and 300 pc.

\section{conclusions}
We have analyzed four {\it BeppoSAX} observations  of NGC7172 and 
ESO103-G35 (two observations for each galaxy separated 
 by about a year). The goal was to search for spectral variability 
 in a large energy band and therefore 
 to attempt to constrain the geometry of the circumnuclear matter in 
these two galaxies.
 The spectra of both galaxies are fitted by an absorbed power-law 
 plus a neutral Fe line in agreement with previous \ginga and \asca observations. 
 A neutral Fe edge is probably detected in the 1996 observation of NGC7172. 
 Our data do not require a reflection component at a statistically significant level.    
 We detected flux variability in both objects by about a factor of about two. 
 We find no significant evidence for spectral variability in the sense 
 that  the power-law spectral index and the column density have remained constant 
 in both galaxies implying a homogeneous obscuring screen and 
 constant physical conditions in the accretion disk corona despite the
 continuum flux variations 
(but see Warwick et al. 1988, Georgantopoulos \& Papadakis 2000). 
 However, in the case of NGC7172 we detect a variation of the 
 Fe line flux by a factor of two following the power-law flux,
 confirming previous \asca results. 
 This provides strong support to a scenario where the Fe line 
  at least in the case of NGC7172 originates in a region close to the nucleus.
 These findings support the idea that one region responsible for the line emission
should lie within one light year away from the nucleus.
This region may be associated with the outer part of an accretion disk. 

The above results clearly emphasize the 
 strength of monitoring observations in probing the 
 geometry of the central engine in AGN. 
 High resolution spectroscopic observations with gratings with 
 both Chandra and XMM are expected to revolutionize our knowledge of the 
 physical conditions of the circumnuclear matter in AGN. 
 These combined with future monitoring observations 
 with large effective area instruments such as EPIC on-board XMM 
 are expected to provide for the first time a detailed mapping 
 of the central regions in AGN. 

\section{acknowledgements}
We thank the anonymous referee for  many useful comments and suggestions.  
This project was funded by a Greek-Italian scientific collaboration 
 under the title ``Observations of 
 active galaxies with the Italian astrophysics mission {\it BeppoSAX}''. AA is grateful  
to the Bologna observatory group for their warm hospitality. 
 We are grateful to M. Cappi, C.  Vignali and R. Lazio for many useful 
  discussions.

\end{document}